\documentclass[preprintnumbers,amsmath,amssymb]{revtex4}
\usepackage{graphicx}
\usepackage{dcolumn}
\usepackage{bm}
\begin{document}
\title{Fokker-Planck equation approach to vehicle statistics}
\author{Dirk Helbing}
  \email{helbing@trafficforum.org}
  \homepage{http://www.helbing.org}
\author{Martin Treiber}
  \email{martin@mtreiber.de}
  \homepage{http://www.mtreiber.de}

\affiliation{Institute for Economics and Traffic, 
  Dresden University of Technology,
  Andreas-Schubert-Str. 23, 01062 Dresden, Germany
}
\date{\today}
\begin{abstract}
This contribution presents a derivation of the steady-state distribution of velocities and
distances of vehicles in freeway traffic which has been suggested for the evaluation of 
interaction potentials among vehicles (see preprint cond-mat/0301484). Despite the
forwardly directed interactions and the additional driving terms in vehicle traffic,
the steady-state velocity and distance distributions agree 
with the equilibrium distributions of classical many-particle systems with
symmetrical interactions, if the system is large enough. Finally, this analytical result is confirmed by
computer simulations.
\end{abstract}

\maketitle

In the particular driven-many particle system we discuss, driver-vehicle
units play the role of the particles. Here, we will describe their behavior
by the coupled car-following equations
\begin{equation}
 \frac{dv_i}{dt} = \frac{v_0 - v_i}{\tau} + f(s_i) - \gamma f(s_{i-1}) + \xi_i(t) \, ,
\label{OV}
\end{equation}
where $v_i(t) = dr_i/dt$ is the speed of vehicle $i$ at time $t$, $v_0$ the maximum velocity,
$s_i(t) = r_{i}(t) - r_{i+1}(t)$ the distance, and $\xi_i(t)$ represents a white noise fluctuation term.
The term $\gamma f(s_{i-1})$ with $0 \le \gamma \le 1$ allows to study different cases:
$\gamma = 0$ corresponds to the case of forwardly directed interactions of vehicles, while
$\gamma = 1$ corresponds to symmetrical interactions of classical particles
in forward and backward direction fulfilling the physical law of ``actio~= reactio''. 

The above stochastic differential equation (Langevin equation)
can be rewritten in terms of an equivalent Fokker-Planck equation. With the definitions
\begin{eqnarray}
 W(s_i) &=& v_0 + \tau [f(s_i) - \gamma f(s_{i-1})] \, , \nonumber \\
 f(s_i) &=& -  \frac{\partial U(s_i)}{\partial s_i} \, , \nonumber \\
 \langle \xi_i(t) \rangle &=& 0 \, , \nonumber \\
 \langle \xi_i(t) \xi_j(t') \rangle &=& D \delta_{ij} \delta (t-t') \, , \nonumber \\
 \mbox{and } P &=& P(s_1,\dots,s_n,v_1,\dots,v_n,t) \, , 
\end{eqnarray}
this Fokker-Planck equation reads
\begin{equation}
 \frac{\partial P}{\partial t} 
 =  \sum_{i=1}^n \bigg\{ - \frac{\partial}{\partial s_i} [ \underbrace{(v_{i} - v_{i+1})}_{=ds_i/dt} P]
 - \frac{\partial}{\partial v_i} \left[\left( \frac{W(s_i) - v_i}{\tau} \right) P \right] 
 + \frac{D}{2} \frac{\partial^2 P}{\partial v_i{}^2} \bigg\} \, , 
\label{FPG}
\end{equation}
where we assume periodic boundary conditions $v_{k+n}(t) = v_k(t)$ and $s_{k+n}(t) = s_k(t)$
for a freeway of length $L$. In the following, we will show that the {\em ansatz} 
\begin{equation}
 P(s_1,\dots,s_n,v_1,\dots,v_n) =
 {\cal N} \mbox{e}^{-\sum_j [U(s_j)/\theta + B s_j]} 
 \mbox{e}^{-\sum_j (v_j - V)^2/(2\theta)} 
\label{distr}
\end{equation}
is a stationary solution of the above Fokker-Planck equation, if the parameters $V$ and $\theta$
are properly chosen. The parameter $B$ is required to specify the actual vehicle density 
(i.e. to ensure $\sum_j s_i = L$). 
\par
In Eq.~(\ref{distr}), 
\begin{equation}
 {\cal N} = \left[ \int ds_1 \dots \int ds_n \int dv_1 \dots \int dv_n \;
  \mbox{e}^{-\sum_j [U(s_j)/\theta + B s_j]} 
 \mbox{e}^{-\sum_j (v_j - V)^2/(2\theta)} \right]^{-1}
\end{equation}
is the normalization constant,
\begin{equation}
 V(t) = \langle v_i \rangle = \int ds_1 \dots \int ds_n \int dv_1 \dots \int dv_n \;
 v_i P(s_1,\dots,s_n,v_1,\dots,v_n,t)
\end{equation}
is the average vehicle velocity, and
\begin{equation}
 \theta(t) = \langle (v_i -V)^2 \rangle = \int ds_1 \dots \int ds_n \int dv_1 \dots \int dv_n \;
 (v_i - V)^2 P(s_1,\dots,s_n,v_1,\dots,v_n,t)
\end{equation}
the velocity variance. In the following, we will restrict our investigation to the stationary case
with $dV/dt = 0$ and $d\theta/dt = 0$, which presupposes that the instability condition of
Eq.~(\ref{OV}) is not fulfilled. For traffic systems with 
$\gamma = 0$, the instability condition is known to be of the form
\begin{equation}
 \frac{dW(s)}{ds} > \frac{1}{2\tau} \, .
\end{equation}
If this condition applies, stop-and-go traffic will emerge.
\par
Differentiation of (\ref{distr}) gives:
\begin{equation}
 - \sum_i \frac{\partial}{\partial s_i} [(v_{i} - v_{i+1})P] 
= \sum_i (v_{i} - v_{i+1} ) \left[ \frac{1}{\theta} \frac{\partial U(s_i)}{\partial s_i} + B \right] P 
= \sum_i (v_{i} - v_{i+1}) \left[B - \frac{f(s_i)}{\theta} \right] P \, ,
\end{equation}
\begin{equation}
 - \sum_i \frac{\partial}{\partial v_i} \left( \frac{W(s_i) - v_i}{\tau} P \right) 
 = \sum_i \frac{P}{\tau} + \sum_i \frac{v_i - W(s_i)}{\tau} \left[ - \frac{(v_i - V)}{\theta}\right] P
\end{equation}
and
\begin{equation}
 \sum_i \frac{D}{2} \frac{\partial^2 P}{\partial v_i{}^2} 
= \frac{D}{2} \sum_i \left[ - \frac{1}{\theta} 
+ \left( - \frac{v_i-V}{\theta} \right)^2 \right] P \, .
\end{equation}
We will now insert this into Eq.~(\ref{FPG}) and 
use the fact that
\begin{equation}
 \sum_i g_{i\pm 1}P = \sum_i g_i P
\end{equation}
for any $i$-dependent variable $g_i$, i.e. indices can be shifted because of the assumed periodic boundary
conditions. In this way 
we find
\begin{eqnarray}
 \frac{\partial P}{\partial t} &=& - \frac{1}{\theta} \sum_i (v_{i} - v_{i+1}) f(s_i) P + \sum_i \frac{P}{\tau}
- \sum_i \frac{DP}{2\theta}  \nonumber \\
&+& \frac{1}{\theta} \sum_i 
\left[ \frac{v_0 - v_i}{\tau} + f(s_i) - \gamma f(s_{i-1})\right] (v_i - V)  P + \frac{D}{2 \theta^2} \sum_i  (v_i - V)^2 P \, .
\end{eqnarray}
Ansatz~(\ref{distr}) can only be a stationary solution with $\partial P/\partial t = 0$,  if
\begin{equation}
 \frac{1}{\theta} = \frac{2}{D\tau} \, ,
\end{equation}
which relates to the fluctuation-dissipation theorem. With this, 
$(v_{i+1}-v_i) = (v_{i+1} - V) - (v_i - V)$, and $(v_0 - v_i) = (v_0 - V) - (v_i - V)]$, we find
\begin{eqnarray}
 \frac{\partial P}{\partial t} = \sum_i 
\frac{1- \gamma}{\theta}  (v_{i+1}-V) 
f(s_i) P  + \frac{1}{\theta} \sum_i \frac{(v_0 - V)(v_i - V)}{\tau}  P \, .
\label{compare}
\end{eqnarray}
We will distinguish the following cases:
\begin{itemize}
\item[1.] In the case of a classical many-particle system with momentum conservation ($\gamma = 1$) 
and energy conservation, i.e. no driving ($v_0 = 0$) and no dissipation ($\tau \rightarrow \infty$), 
we find $\partial P/\partial t = 0$,
i.e. ansatz (\ref{distr}) is an exact stationary solution of the Fokker-Planck equation (\ref{FPG}).
\item[2.] In the case of vehicle traffic ($\gamma = 0$), we have to show that the additional term 
\begin{equation}
 \frac{1}{\theta} \sum_i (v_{i+1}-V) \left[ f(s_i)   + \frac{v_0 - V}{\tau}\right]  P 
\label{disap}
\end{equation}
disappears (where we have again shifted indices). 
Let us first note that, with the factorization assumption, we can state
\begin{equation}
 \lim_{n\rightarrow \infty} \frac{1}{n} \sum_i (v_{i+1}-V) \left[ f(s_i)   + \frac{v_0 - V}{\tau}\right]
= \left[ \lim_{n\rightarrow \infty} \frac{1}{n} \sum_i (v_{i+1}-V) \right]
\left\{ \lim_{n\rightarrow \infty} \frac{1}{n} \sum_i \left[ f(s_i)   + \frac{v_0 - V}{\tau}\right] \right\} \, . 
\end{equation}
The first factor vanishes because of $V = \lim_{n\rightarrow \infty} \frac{1}{n} \sum_i v_i$,
but the second factor disappears as well:
Dividing Eq.~(\ref{OV}) by $n$ and summing up over $i$ gives
\begin{equation}
 \frac{1}{n} \sum_i \frac{dv_i}{dt} = \frac{1}{n} \sum_i \frac{v_0 - v_i}{\tau}
+ \frac{1}{n} \sum_i f(s_i) + \frac{1}{n} \sum_i \xi_i(t) \, . 
\end{equation}
In the limit $n\rightarrow \infty$ of large enough particle numbers $n$, 
the left-hand side converges to $dV/dt$, while the last term on the right-hand side converges
to 0. In the assumed stationary case with $dV/dt = 0$ and using
$v_0 - v_i = (v_0 - V) - (v_i - V)$, this implies
\begin{equation}
 \lim_{n\rightarrow \infty}
 \frac{1}{n} \sum_i \left[ \frac{v_0 - v_i}{\tau} + f(s_i) \right] = 0 \quad \mbox{and} \quad
 \lim_{n\rightarrow \infty} \frac{1}{n} \sum_i \left[ \frac{v_0 - V}{\tau} + f(s_i) \right] = 0 \, . 
\label{because}
\end{equation}
\end{itemize}
Nevertheless, it is not obvious what happens for finite systems, as the standard deviation of
$\sum_i (v_{i+1} - V)$ is $\sqrt{n\theta}$. In order to see how the single-particle distributions depend on 
$n$, let us again assume a factorizing solution 
\begin{equation}
 P(s_1,\dots,s_n,v_1,\dots,v_n,t) = \prod_{i=1}^n g(s_i,t) \prod_{j=1}^n h(v_j,t) 
\label{factor}
\end{equation}
in generalization of ansatz (\ref{distr}).
Inserting this into (\ref{compare}) with $\gamma = 0$ gives
\begin{eqnarray}
 \frac{\partial P}{\partial t} = \sum_i \frac{\frac{d}{dt} [g(s_i,t)h(v_{i+1},t)]}{g(s_i,t)h(v_{i+1},t)}P
= \frac{1}{\theta} \sum_i (v_{i+1}-V) \left[ f(s_i)   + \frac{v_0 - V}{\tau}\right]  P 
\end{eqnarray}
or
\begin{eqnarray}
 \frac{1}{n} \sum_i \frac{d}{dt} [g(s_i,t)h(v_{i+1},t)]
&=&  \frac{1}{n \theta} \sum_i (v_{i+1}-V) \left[ f(s_i)   + \frac{v_0 - V}{\tau}\right]  g(s_i,t)h(v_{i+1},t) \, . 
\end{eqnarray}
The left-hand side represents the average temporal change of the one-particle distribution functions
$g(s_i,t)$ of the netto distance and $h(v_{i+1},t)$ of the speed, while the right-hand side converges to
zero with growing system size according to the central limit theorem and the factorization
assumed with Eq.~(\ref{factor}), i.e. the statistical independence of the variables $v_i$ and $s_i$.  
That is, while (\ref{distr}) is an exact equilibrium solution for a classical many-particle system,
it is also expected to be a steady-state solution of driven many-particle systems of 
the kind (\ref{OV}), even if the potential is forwardly directed rather than symmetric!
\par
\begin{figure}[htbp]
\includegraphics[width=85mm]{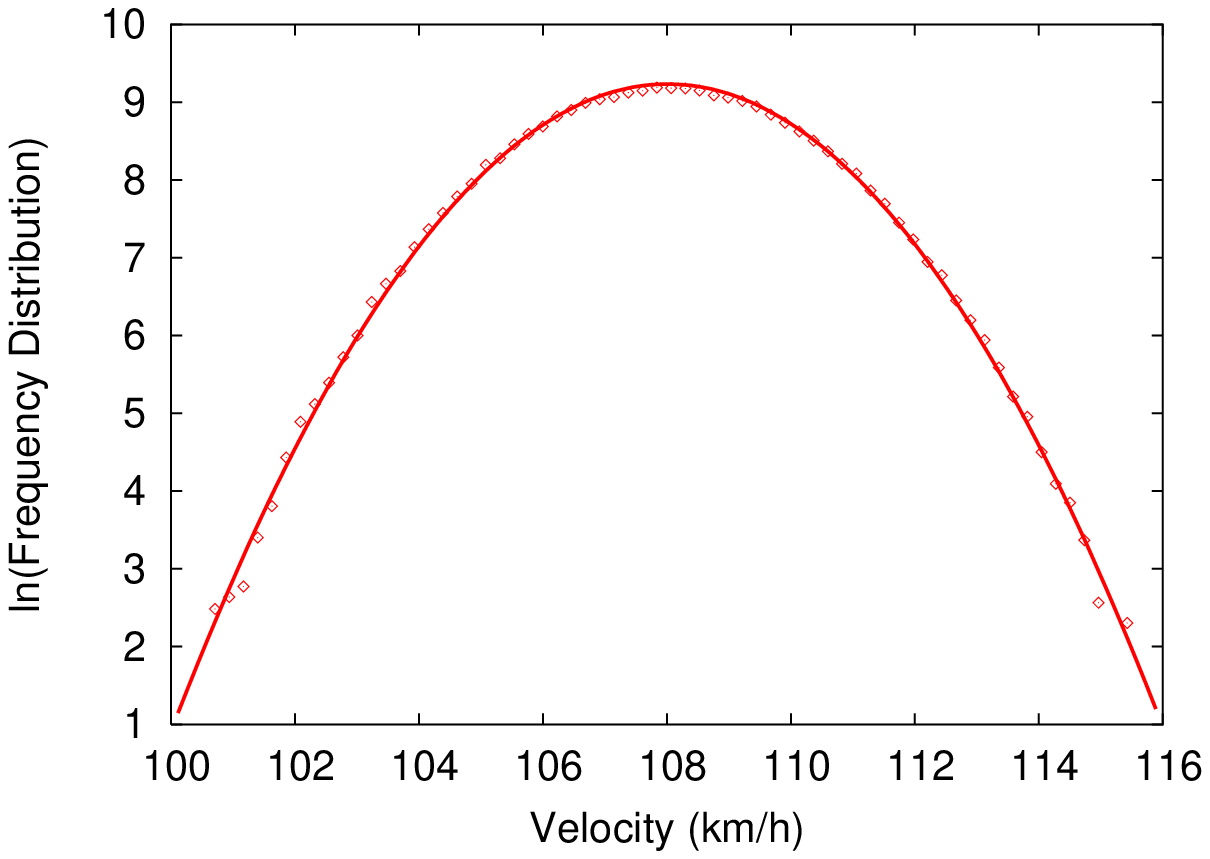}
\includegraphics[width=85mm]{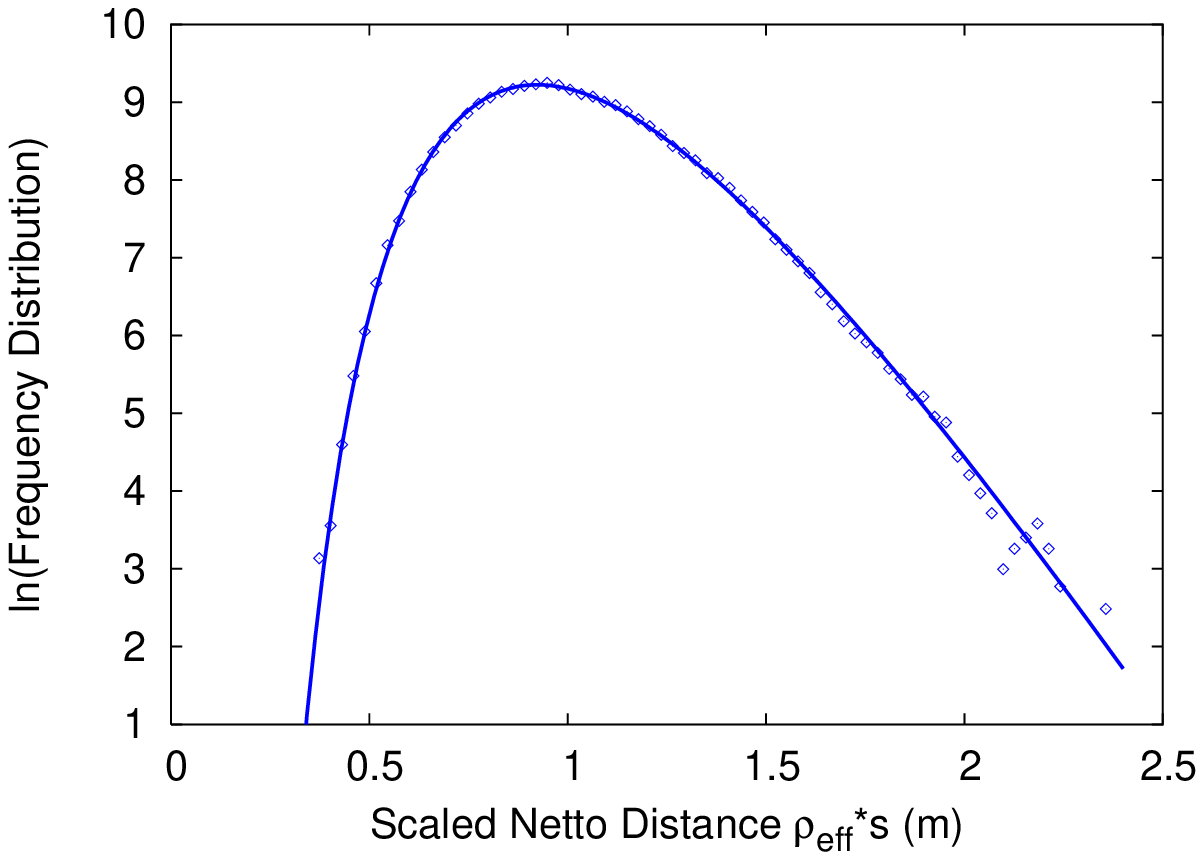}\\
\includegraphics[width=85mm]{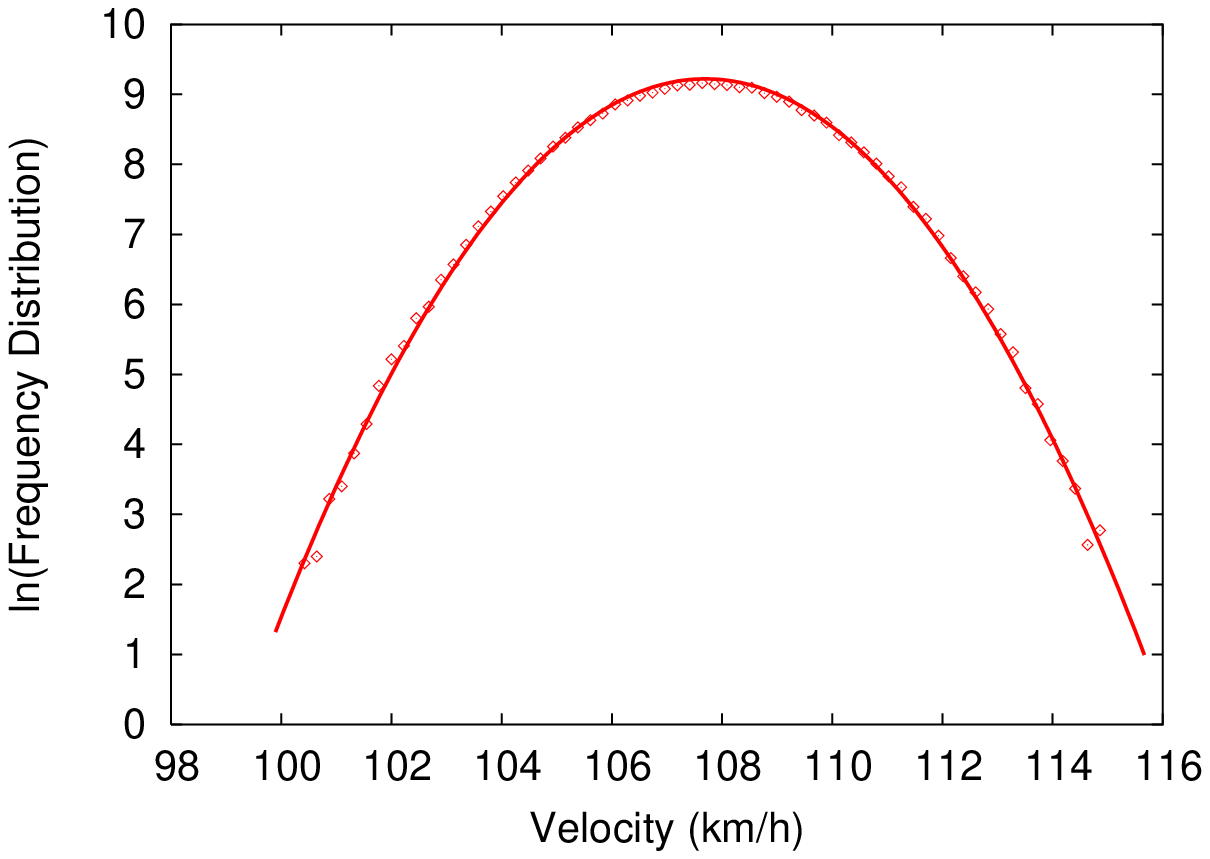}
\includegraphics[width=85mm]{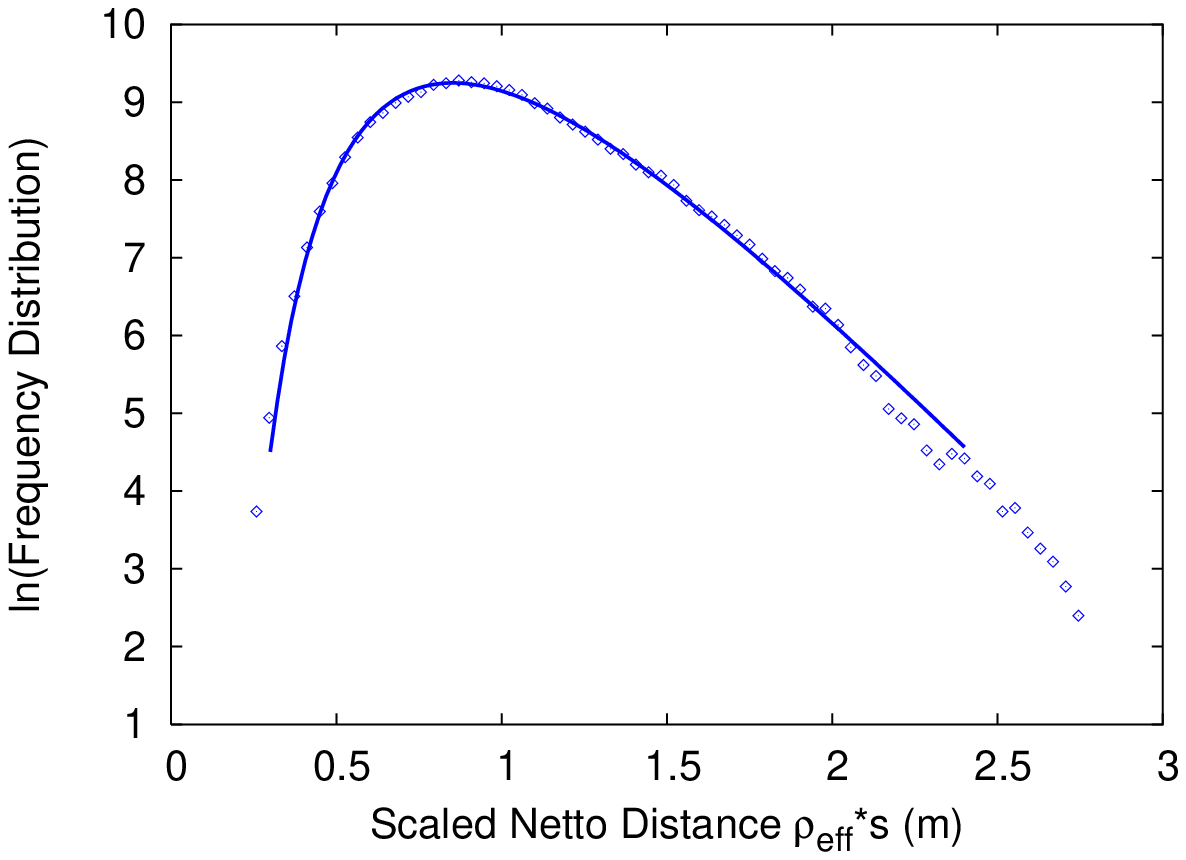}
\caption[]{Semilogarithmic plot of the frequency distributions of the velocity (left) and the 
distance (right) according to numerical simulations for symmetrical
interactions (top) and forwardly-directed ones (below). The assumed interaction potential
was $U(s) \propto s^{-1}$ for $s> 0$ and 0 otherwise.}
\label{Fig1}
\end{figure}
\begin{figure}[htbp]
\includegraphics[width=85mm]{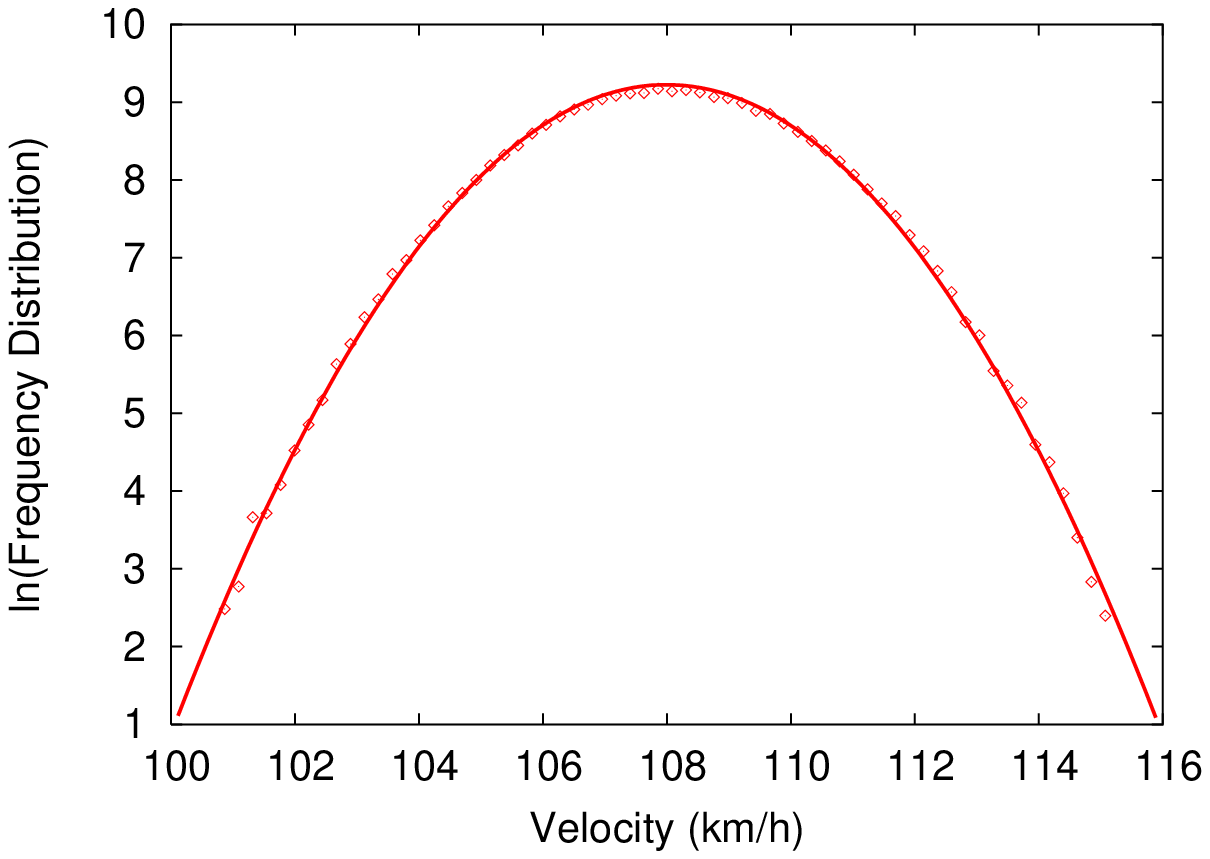}
\includegraphics[width=85mm]{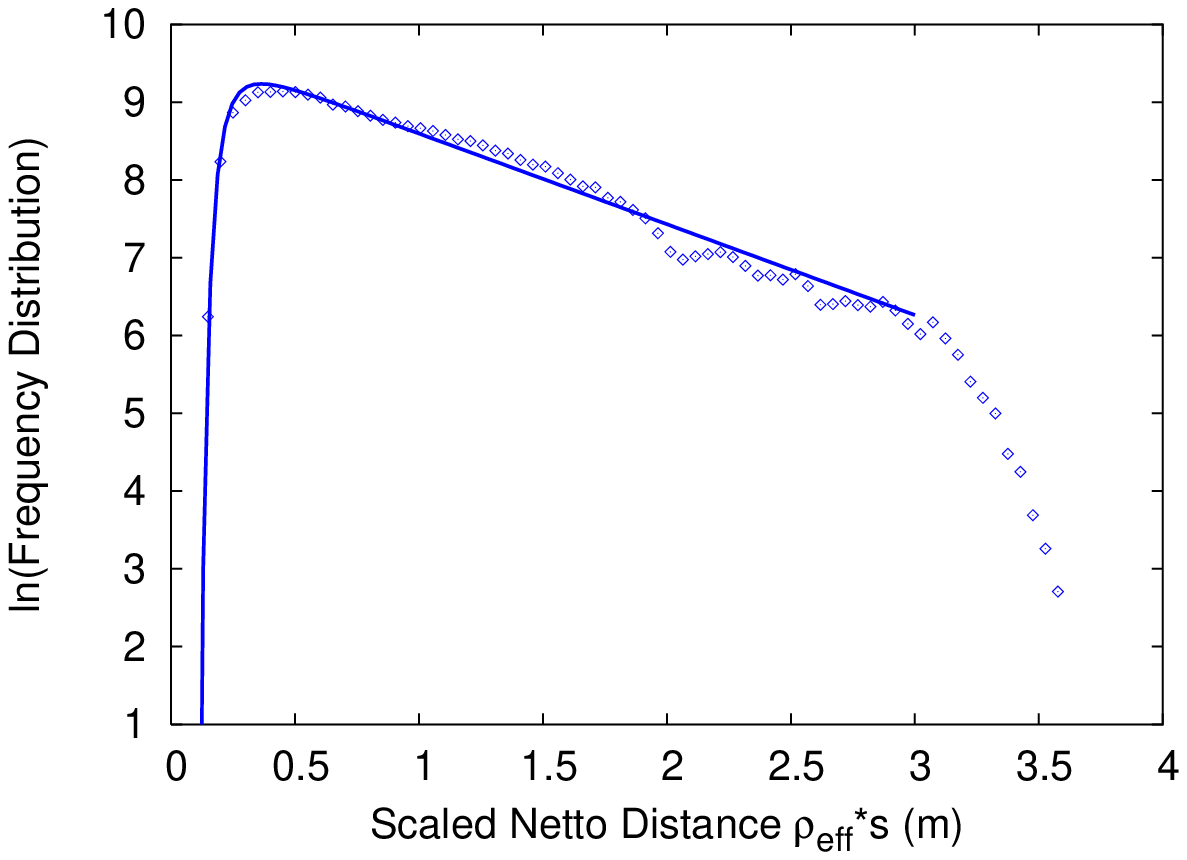}\\
\includegraphics[width=85mm]{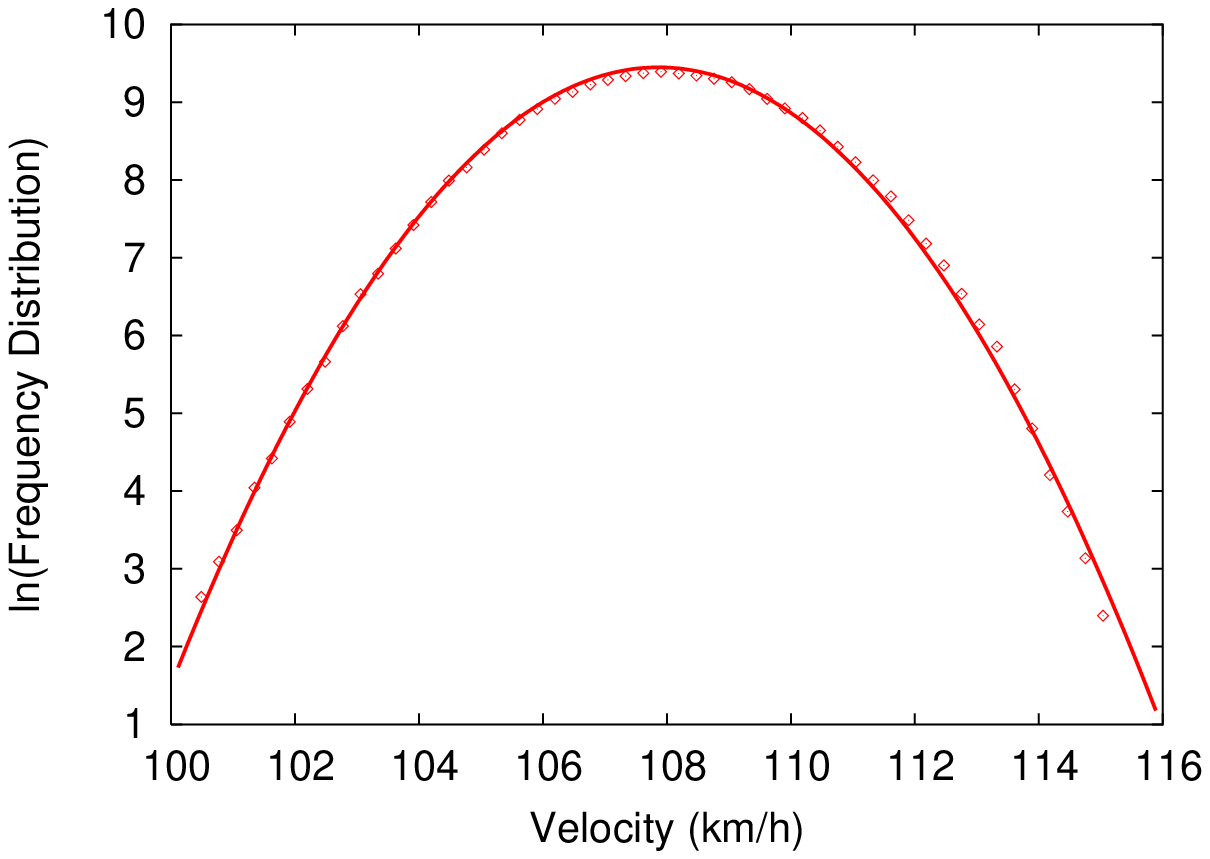}
\includegraphics[width=85mm]{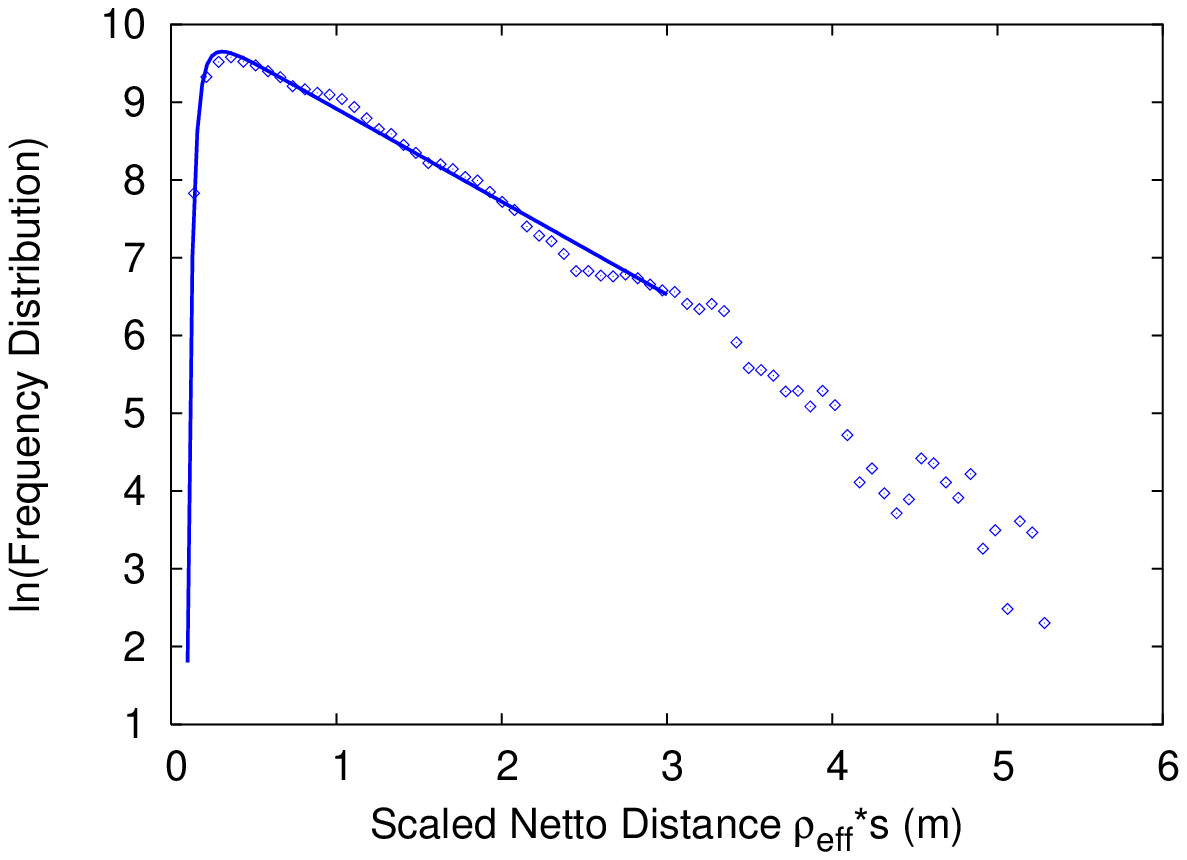}
\caption[]{Semilogarithmic plot of the frequency distributions of the velocity (left) and the 
distance (right) according to numerical simulations for symmetrical
interactions (top) and forwardly-directed ones (below). The applied interaction potential
was $U(s) \propto s^{-4}$ for $s> 0$ and 0 otherwise.}
\label{Fig2}
\end{figure}

Note that, because of the factorization ansatz (\ref{distr}) and (\ref{factor}), one may use the approximation
 \begin{equation}
 \frac{1}{n} \sum_i (v_{i+1}-V) \left[ f(s_i)   + \frac{v_0 - V}{\tau}\right]  
\approx \left[ \frac{1}{n} \sum_i (v_{i+1}-V) \right]
 \left\{ \frac{1}{n} \sum_i \left[ f(s_i)   + \frac{v_0 - V}{\tau}\right] \right\} \, , 
\end{equation}
where {\em both} factors on the right-hand side converge to zero because of (\ref{because}). 
Therefore, the convergence
should be particularly fast. Moreover, in empirical evaluations, the estimator of $V$ is
$\frac{1}{n} \sum_i v_i$, i.e. the first factor on the right-hand becomes exactly zero. For all these reasons,
it is expected that 
\begin{equation}
 g(s) \propto \mbox{e}^{-[U(s)/\theta + B s]} 
\end{equation}
is a good approximation of the empirical distance distribution
and 
\begin{equation}
h(v)\propto \mbox{e}^{-(v - V)^2/(2\theta)}
\end{equation}
a good approximation of the empirical velocity distribution,
where $V = \frac{1}{n} \sum_i v_i$, $\theta = \frac{1}{n-1} \sum_i (v_i - V)^2$, and $n > 50$.
This is actually confirmed by numerical simulations of Eq.~(\ref{OV}), see Figs.~\ref{Fig1} and
\ref{Fig2}. The numerical results (symbols) agree with the steady-state solution (solid line).
Within the statistical variation and apart from finite size corrections, 
the results are the same for symmetrical and forwardly directed interaction
potentials.
\par
Finally, let us investigate the Hamiltonian
\begin{equation}
 {\cal H} = {\cal T} + {\cal V}
 = \sum_i \frac{(v_i - V)^2}{2} + \sum_i U(s_i) \, .
\end{equation}
If $dV/dt = 0$, we can derive the following relations:
\begin{eqnarray}
\frac{d{\cal H}}{dt} &=& \frac{d{\cal T}}{dt} + \frac{d{\cal V}}{dt} \nonumber \\
&=& \sum_i (v_i - V) \frac{dv_i}{dt}  + \sum_i \frac{\partial U(s_i)}{\partial s_i} 
\left(\frac{ds_i}{dr_{i}}\frac{dr_{i}}{dt} + \frac{ds_{i}}{dr_{i+1}} \frac{dr_{i+1}}{dt} \right) \nonumber \\
&=& \sum_i (v_i - V) \frac{dv_i}{dt}  - \sum_i f(s_i) (v_{i} - v_{i+1}) \nonumber \\
&=& \sum_i (v_i - V) \left(  \frac{v_0 - v_i}{\tau} + f(s_i) - \gamma f(s_{i-1}) + \xi_i(t) \right) 
+ \sum_i f(s_i) (v_{i+1} - v_{i}) \nonumber \\
&=& \sum_i (1-\gamma) (v_{i+1}-V) f(s_i)  + \sum_i \frac{(v_0 - V)(v_i - V)}{\tau} \nonumber \\
&-& \sum_i \frac{(v_i - V)^2}{\tau} + \sum_i (v_i - V) \xi_i(t) \, .
\end{eqnarray}
Comparing this with (\ref{compare}) shows that 
\begin{equation}
 \frac{\partial P}{\partial t} = \frac{P}{\theta} \frac{d{\cal H}}{dt} +\frac{1}{\theta} \sum_i \left[
   \frac{(v_i - V)^2}{\tau} - (v_i - V) \xi_i(t) \right] P \, .
\end{equation}
Correspondingly, we have
\begin{equation}
 \frac{d{\cal H}}{dt} = \sum_i \left[ (v_i - V) \xi_i(t) - \frac{(v_i-V)^2}{\tau} \right] 
= \sum_i (v_i -V)\left( \xi_i(t) - \frac{v_i - V}{\tau} \right) 
\label{H}
\end{equation}
in the stationary state $\partial P/\partial t = 0$. We will again distinguish two different cases:
\begin{itemize}
\item[1.] In a conservative system with no fluctuations ($\xi_i(t) = 0 = D$) and no
dissipation ($\tau \rightarrow \infty$), we have $d{\cal H}/dt = 0$, independently of whether
the interactions are symmetric or forwardly directed.
\item[2.] For many-particle systems with fluctuation terms and/or dissipation, one
can show
\begin{eqnarray}
 \langle \xi_i (v_i - V) \rangle &=& \left\langle \frac{1}{2} \frac{d(v_i - V)^2}{dt} \right\rangle
- \frac{v_0-V}{\tau} \sum_i \langle v_i - V \rangle + \frac{1}{\tau} \langle (v_i -V)^2 \rangle
- \langle [f(s_i) - \gamma f(s_{i-1})] (v_i-V) \rangle \nonumber \\
 &=& \frac{1}{2} \frac{d\theta}{dt} - \frac{v_0-V}{\tau}(\langle v_i \rangle - V)  + \frac{\theta}{\tau} 
- \langle f(s_i) - \gamma f(s_{i-1}) \rangle (\langle v_i \rangle -V ) \, .
\label{show}
\end{eqnarray}
This can be found by multiplication of Eq.~(\ref{OV}) with $(v_i-V)$ and calculation of the 
ensemble average, using the factorization ansatz (\ref{distr}) or (\ref{factor}).
The first term on the right-hand side vanishes under the assumption of a stationary state.
The second and the fourth term vanish because of $\langle v_i \rangle = V$. Therefore,
\begin{equation}
 \langle \xi_i (v_i - V) \rangle = \frac{\theta}{\tau} 
\qquad \mbox{and} \qquad \frac{1}{n} \sum_i \xi_i (v_i - V) \approx  \frac{1}{n} \sum_i \frac{(v_i -V)^2}{\tau} \, ,
\end{equation}
if $n$ is large enough.
Without dissipation ($\tau \rightarrow \infty$), $\langle \xi_i (v_i - V) \rangle$ becomes zero,
while it is finite otherwise. Together with Eq.~(\ref{H}), we arrive at 
\begin{equation}
 \left\langle \frac{d{\cal H}}{dt} \right\rangle = 0 \, ,
\end{equation}
i.e. in the statistical average we have $d{\cal H}/dt = 0$. This is also expected for systems with many
particles. 
\end{itemize}
In conclusion, the equilibrium solution (\ref{distr}) of conservative many-particle systems is a
good approximation for steady-state solutions ($\partial P/\partial t$) of driven many-particle systems of the
kind (\ref{OV}) with asymmetrical interactions, driving and dissipation effects, if the system is
large enough, i.e. $n\gg 1$. For small systems, we expect that fluctuations become essential. 
A more detailed elaboration is presently in preparation and will be submitted, soon.
\end{document}